\documentclass[aps,prd,preprint,showpacs,nofootinbib]{revtex4}
\usepackage{amsmath,amssymb}
\usepackage[dvipdfmx]{graphicx}
\usepackage{color}
\usepackage{comment}

\def\mydate{23 May 2012}
\def\ignore#1{{}}

\ignore{

\makeatletter
\@addtoreset{equation}{section}
\makeatother
}

\newcommand{\simg}{%
\hspace{0.3em}\raisebox{0.4ex}{$>$}\hspace{-0.75em}\raisebox{-.7ex}{$\sim$}\hspace{0.3em}} 

\newcommand{\siml}{%
\hspace{0.3em}\raisebox{0.4ex}{$<$}\hspace{-0.75em}\raisebox{-.7ex}{$\sim$}\hspace{0.3em}}

\newcommand{\beeq}{\begin{equation}}
\newcommand{\eneq}{\end{equation}}
\newcommand{\beqn}{\begin{eqnarray}}
\newcommand{\eeqn}{\end{eqnarray}}

\def\la{\raise.16ex\hbox{$\langle$}\lower.16ex\hbox{}  }
\def\ra{\raise.16ex\hbox{$\rangle$}\lower.16ex\hbox{} }
\def\go{\rightarrow}

\def\onehalf{ \hbox{$\frac{1}{2}$} }

\def\eff{{\rm eff}}

\def\KK{{\rm KK}}

\def\mbar{\overline{m}}
\def\Qbar{\overline{Q}}

\def\ep{\epsilon}
\def\psibar{ \psi \kern-.65em\raise.6em\hbox{$-$} }
\def\psibarl{ \psi \kern-.65em\raise.6em\hbox{$-$} \lower.6em\hbox{} }

\begin{document}


\preprint{OU-HET 728/2011}

\date{\mydate}

\title{SUSY breaking scales in the  gauge-Higgs unification}

\author{Hisaki Hatanaka}
\author{Yutaka Hosotani}


\affiliation{Department of Physics, 
Osaka University, 
Toyonaka, Osaka 560-0043, 
Japan}



\begin{abstract}
In the $SO(5) \times U(1)$ gauge-Higgs unification
in the Randall-Sundrum (RS) warped space the Higgs boson naturally 
becomes stable.  The model is consistent with the current collider signatures
for a large warp factor $z_L > 10^{15}$ of the RS space.
In order for stable Higgs bosons to explain  the dark matter of the  
Universe the Higgs boson must have a mass $m_h = 70 \sim 75 \,$GeV, which
can be obtained in the non-SUSY model with $z_L \sim 10^5$.
We show that this discrepancy is resolved in  supersymmetric gauge-Higgs
unification where a stop mass is about $300 \sim 320 \,$GeV 
and gauginos in the electroweak sector are light.
\end{abstract}


\maketitle



The Higgs boson,  necessary for inducing spontaneous symmetry breaking 
in the standard model (SM) of  electroweak interactions,  is yet to be discovered.
It is not clear at all if the Higgs boson appears as described in the SM.
New physics may be hiding behind it,   the Higgs boson having properties
quite different from those in the SM.

In the gauge-Higgs unification scenario the 4D Higgs boson becomes a part
of the extra-dimensional component of gauge fields.\cite{YH1}-\cite{Scrucca}
Many  models have 
been proposed with predictions to be tested at colliders.  
Among them the $SO(5) \times U(1)$ gauge-Higgs unification in the 
Randall-Sundrum (RS)  warped space is  most promising.\cite{ACP}-\cite{Contino2011}

One of the most striking results in the model is that the 4D Higgs boson 
naturally becomes stable.\cite{HKT}  The vacuum expectation value of the 
Higgs boson corresponds to an Aharonov-Bohm (AB) phase $\theta_H$ 
in the fifth dimension.  With bulk fermions introduced in the vector
representation of $SO(5)$ the value of $\theta_H$ is dynamically
determined to be $\onehalf \pi$, at which the Higgs boson becomes
stable while giving masses to quarks, leptons, and weak bosons.
There emerges $H$ parity ($P_H$) invariance at $\theta_H = \onehalf \pi$.
All particles in the SM other than the Higgs boson are $P_H$-even, while
the only $P_H$-odd particle at low energies is the Higgs boson, which
in turn guarantees the stability of the Higgs boson.\cite{HTU1, Contino2011}
As a consequence the Higgs boson cannot be seen in the current collider
experiments, since all experiments so far are designed to find decay
products of the Higgs boson.  

The model has one parameter to be determined, namely the warp factor $z_L$ 
of the RS spacetime.  With $z_L$ given, the mass of the Higgs boson $m_h$ is
predicted.  It is found that $m_h = 72, 108$ and 135$\,$GeV for
$z_L = 10^5, 10^{10}$ and $10^{15}$, respectively.  We note that the LEP2
bound,  $m_h > 114\,$GeV,  is evaded as the $ZZH$ coupling
exactly vanishes as a result of the $P_H$ invariance.

There appears slight deviation in the gauge couplings of quarks and leptons 
from those in the SM.  It turns out that the gauge-Higgs unification model gives 
a better fit to the forward-backward asymmetries in $e^+ e^-$ collisions on 
the $Z$ pole than the SM.  However, the branching fractions of $Z$ decay are 
fit well only for $z_L \simg 10^{15}$.  The gauge-Higgs unification model
gives  predictions for Kaluza-Klein (KK) excitation modes of various particles.  
In particular, the first KK $Z$ has a mass 1130$\,$GeV and a width 422$\,$GeV
for $z_L = 10^{15}$.  The current limit on the $Z'$ production at the Tevatron
and LHC indicates $z_L > 10^{15}$.  All of the collider data prefer a large
warp factor in the gauge-Higgs unification model.\cite{HTU2}
These analyses have been done at the tree level so far.

The fact that Higgs bosons become stable leads to another important
consequence.  They become the dark matter of the Universe.\cite{HKT}
It has been shown that
in order for stable Higgs bosons to account for the entire dark matter
of the Universe observed by WMAP,\cite{DM2}   
$m_h$ must be  in the range  $70 \sim 75\,$GeV, 
smaller than the $W$ boson mass $m_W$.  If $m_h > m_W$, the relic
abundance of Higgs bosons would become very small.  
To have  $m_h= 70 \sim 75\,$GeV in the gauge-Higgs unification model
we need $z_L \sim 10^5$, which is in conflict with the collider data.

Of course nothing is wrong with $m_h \sim 135\,$GeV.  It simply implies
that Higgs bosons account for a tiny fraction of the dark matter of the Universe.  
Yet it is curious and fruitful to ask if there is a natural way in the gauge-Higgs
unification scenario to satisfy the two requirements; (i) to be consistent with
the collider data, and (ii) to explain the entire dark matter of the Universe.

In this paper we would like to show that the two requirements are naturally
fulfilled  if the model has softly broken supersymmetry (SUSY) such that
SUSY partners of observed particles acquiring large masses.  It will be
found that the mass of a stop ($\tilde t$), a SUSY partner of a top quark ($t$), 
needs to be  $300\sim 320\,$GeV,
when  SUSY partners of $W$, $Z$ and $\gamma$ are light.

The key observation is that the nonvanishing Higgs boson mass $m_h$ 
in the gauge-Higgs unification arises  at the quantum-level,  whereas the dominant 
part of collider experiments is governed by the structure  at the tree-level.
If SUSY is exact, the contributions of bosons and fermions to the effective
potential  $V_\eff (\theta_H)$ cancel so that $V_\eff (\theta_H)=0$, 
the Higgs boson remaining  massless.   As SUSY is broken, the cancellation
becomes incomplete.  If the SUSY breaking scale is much larger than the KK
mass scale, the model is reduced to the non-SUSY model.  In particular,
$m_h$ becomes $\sim 135\,$GeV for $z_L = 10^{15}$.
Put differently, one can ask how large the SUSY breaking scale should be
to have $m_h = 70 \sim 75\,$GeV with $z_L = 10^{15} \sim 10^{17}$ so that 
the relic abundance of Higgs bosons  saturate the dark matter of the Universe.  

The RS warped spacetime is given by 
$ds^2 = e^{-2ky} dx_\mu dx^\mu + dy^2$
for $0 \le y \le L$.\cite{RS}  The AdS curvature in $0 < y < L$ is $-6 k^2$.
The warp factor is $z_L = e^{kL}$.  
In the $SO(5) \times U(1)$ gauge-Higgs unification 
there appears an AB phase,  or the Wilson line phase $\theta_H$,  in the 
fifth dimension, as the RS spacetime has topology of $R^4 \times (S^1/Z_2)$.
The 4D Higgs field appears as a zero mode in the $SO(5)/SO(4)$ part of the 
fifth-dimensional component  of the vector potential $A_y (x,y)$.  
$\la A_y \ra  = \la A_y^{\hat 4} \ra T^{\hat 4}  \not= 0$  
when the EW symmetry is spontaneously broken, where  
the $SO(5)$ generator $T^{\hat 4}$ is defined by
$(T^{\hat 4})_{ab} = (i/\sqrt{2})(\delta_{a5} \delta_{b4} -\delta_{a4} \delta_{b5})$.
The Wilson line phase $\theta_H$ is given by
$\exp \big\{  i \theta_H \sqrt{2} T^{\hat 4} \big\} 
= P \exp \big\{ ig_A \int_0^L dy \la A_y \ra \big\}$.

The effective potential $V_\eff (\theta_H)$ at the one-loop level is determined
by the mass spectrum $\{ m_n (\theta_H) \}$ in the presence of the  phase 
$\theta_H \not= 0$. It is given in $d$ dimensions, after Wick rotation, by
\beqn
&&\hskip -1.cm
V_\eff = \pm  \frac{1}{2} \int \frac{d^d p}{(2\pi)^d} 
\sum_n \ln (p^2 + m_n^2)  \cr
\noalign{\kern 5pt}
&&\hskip -.3cm
=  \mp \frac{\Gamma ( - \onehalf d)}{2 (4\pi)^{d/2}} \sum_n m_n^d  ~.
\label{effV1}
\eeqn
The upper (lower) sign corresponds to bosons (fermions).  The second equality is 
understood by analytic continuation for large ${\rm Re} \, d$ in the complex $d$-plane.

In supersymmetric theory with SUSY breaking each KK tower with a spectrum 
$\{ m_n \}$  is accompanied  by its SUSY partner with a spectrum 
$\{ \mbar_n \}$.   With a SUSY breaking scale $\Lambda$ the latter is 
well mimicked by
\beeq
\mbar_n = \sqrt{ m_n^2 + \Lambda^2 }
\label{spectrum1}
\eneq
which has a property that $\mbar_n \sim m_n$ for $m_n \gg \Lambda$.
Suppose that the  spectrum $\{ m_n \}$ ($m_n >0$) is determined by the zeros of
an analytic function $\rho (z)$; $\rho (m_n) = 0$.   Then 
the spectrum $\{ \mbar_n \}$ is determined by the zeros of
$\bar \rho (z)= \rho(\sqrt{z^2 - \Lambda^2})$.
When $\rho(iy) = \rho(-iy)$ for real $y$  and $|\ln \rho \, | < |z|^q$ with some $q$ 
for $|z| \go \infty$, a convenient formula for $V_\eff$ in (\ref{effV1})  has been
derived.\cite{Garriga} 
In the present case the function $\bar \rho (z)$ has a branch cut between 
$\Lambda$ and $-\Lambda$ in the $z$-plane so that elaboration of
the argument there is necessary.

In  the dimensional regularization,   (\ref{effV1})  
with    $\{ \mbar_n \}$  is transformed into
\beeq
V_\eff = \pm \frac{\Gamma ( 1- \onehalf d)}{2\pi i  (4\pi)^{d/2}} 
\int_C dz \, z^{d-1}  \ln \bar\rho (z)  ~.
\label{effV2}
\eneq
Here the contour $C$ encircles the zeros of $\bar \rho(z)$, $\{ \mbar_n \}$, 
clockwise.  The contour is transformed, for sufficiently small ${\rm Re} \, d$,  
to $C'$ which runs from $-i\infty$ to $+i \infty$,  avoiding the cut, as shown in
fig.\ 1.  With $C'$ the formula is analytically continued to a larger  ${\rm Re} \,d$.
The contribution coming from an infinitesimally small circle ($C_\ep$) around the 
branch point at $\Lambda$ vanishes.  The integral just below the cut ($C^{\rm cut}_-$)
cancels the one just above the cut ($C^{\rm cut}_+$) as $\rho(iy) = \rho(-iy)$.
The contributions coming from the integrals along the  imaginary axis combine
to give the expression involving an integral 
$\int_0^\infty dy \, y^{d-1} \ln \rho(i \sqrt{y^2 + \Lambda^2})$. 
Combining contributions from the KK tower with a spectrum $\{ m_n \}$ and its SUSY partner with $\{ \mbar_n \}$, one finds that
\beeq
V_\eff = \frac{\pm 1}{ (4\pi)^{d/2} \Gamma(\onehalf d) } 
\int_0^\infty dy \, y^{d-1}  \ln \frac{\rho ( iy) }{\rho(i \sqrt{y^2 + \Lambda^2})} ~.
\label{effV3}
\eneq
Previously $V_\eff (\theta_H)$ in the gauge-Higgs unification in the RS spacetime
has been evaluated by making use of the formula with only
$y^{d-1} \ln \rho ( iy)$ in the integrand.\cite{HOOS, HTU1, Garriga, OdaWeiler}

\begin{figure}[htb]
\begin{center}
\includegraphics[width=7cm]{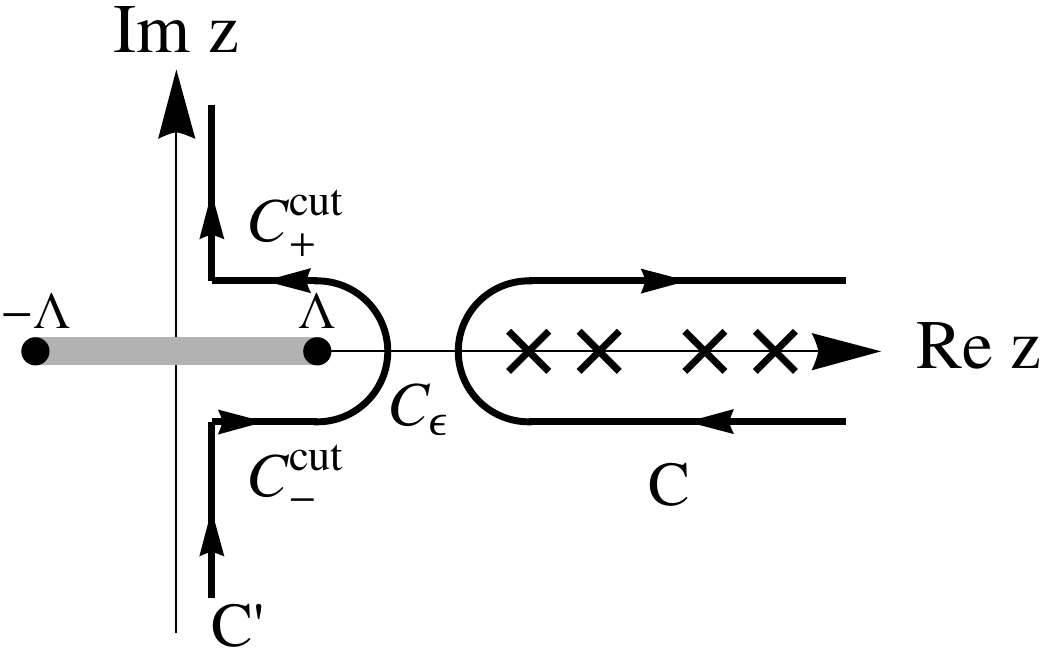}
\caption{Contours $C$ and $C'$ in the expression (\ref{effV2}).}
\label{contour}
\end{center}
\end{figure}

The $SO(5) \times U(1)$ gauge-Higgs unification model has been specified 
in Refs.\ \cite{HOOS, HNU}.  In the bulk five-dimensional spacetime, 
in addition to the $SO(5)$ and $U(1)$ gauge fields, 
four bulk fermion multiplets in the vector representation of $SO(5)$ are
introduced for each generation of quarks and leptons.  
On the Planck brane at $y=0$,  right-handed brane  fermions 
$\hat \chi_R^\alpha$  and one brane scalar $\hat \Phi$ are introduced.
The orbifold boundary conditions break $SO(5)$ to 
$SO(4) \simeq SU(2)_1 \times SU(2)_2$.  
The nonvanishing vacuum expectation value $\la \hat \Phi \ra$
spontaneously breaks $SO(4) \times U(1)$ to $SU(2)_1 \times U(1)_1$, and 
at the same time give large masses of $O(m_\KK)$ to exotic fermions.
The resultant fermion spectrum at low energies is the same as in the SM.
The  $SO(4) \times U(1)$ gauge anomalies are canceled. 

The  relevant contributions to  $V_\eff (\theta_H)$ come from the $W$ 
and $Z$ towers of the four-dimensional components $A_\mu(x,y)$,  
the Nambu-Goldstone towers of the fifth-dimensional  components $A_y (x,y)$, 
and the top quark tower. Contributions coming from other quark and 
lepton towers are negligible.\cite{HOOS, HTU1}
Brane fields give no contribution to $V_\eff (\theta_H)$, as they do not
couple to $A_y$.
We recall  the way the  quarks and leptons acquire masses is 
different from that in SM.  $A_y$ connects the left-handed and right-handed 
components of the up-type quarks and charged leptons directly, whereas 
those of the down-type quarks and neutrinos are intertwined through both gauge couplings
and additional interactions with brane fermions and scalars.  
Thus the effective Higgs couplings of 
the down-type quarks and neutrinos appear after integrating heavy brane fermions.
It is notable that only the ratio of two large mass scales of the brane fermion couplings 
appears in the effective Higgs couplings.\cite{HOOS, HNU}

We comment that in the supersymmetric extension of the model two brane scalar
fields, $\hat \Phi_u$ and $\hat \Phi_d$, need to be introduced. 
Further in 5D SUSY there appear 4D scalar fields,  associated to 
the zero mode of $A_y$, to form a 5D $N=1$ (4D $N=2$) vector multiplet.
In this paper we  assume that such scalar fields acquire large SUSY breaking masses,
giving little effect on the Wilson-line dynamics.

In the supersymmetric extension   two SUSY breaking scales  become
important for $V_\eff (\theta_H)$: 
$\Lambda_{\rm gh}$ for the super partners of the $W$, $Z$ and
Nambu-Goldstone towers, and $\Lambda_{\rm stop}$ for super partner 
of the $t$ quark tower.  The stop ($\tilde t$) mass is given by 
$m_{\tilde t} = \sqrt{ m_t^2 + \Lambda_{\rm stop}^2}$.

There arises no constraint to the masses of other squarks and sleptons.
The masses of gluinos do not affect $V_\eff (\theta_H)$,  being
irrelevant in the present analysis.

It is most convenient to express the function $\rho(z)$ in (\ref{effV3})
in the form $\rho (iy) = 1 + Q(q)$ where $y = k z_L^{-1} q$.  
The  effective potential is expressed in terms of 
\beeq
I[Q, \tilde \Lambda] =
\frac{(kz_L^{-1})^4}{(4\pi )^2} \int_0^\infty dq \, q^3
\ln \frac{1+ Q(q)}{1 + Q[ (q^2 + \tilde \Lambda^2)^{1/2}]}
\eneq
where $\tilde \Lambda$ is related to the SUSY breaking scale by
$\Lambda = k z_L^{-1} \tilde \Lambda$.
We define
\beqn
&&\hskip -1.cm
Q_0 (q; c) = \Qbar_0 (q; c) \sin^2 \theta_H ~, \cr
\noalign{\kern 10pt}
&&\hskip -1.cm
\Qbar_0 (q; c) = \frac{z_L}{q^2 F_{c-(1/2)}(q)  F_{c+(1/2)}(q)} ~, \cr
\noalign{\kern 10pt}
&&\hskip -1.cm
F_\alpha (q) = I_\alpha (qz_L^{-1}) K_\alpha (q) -  K_\alpha (qz_L^{-1}) I_\alpha (q)
\label{func1}
\eeqn
where $I_\alpha (q)$ and $K_\alpha (q)$ are the modified Bessel functions.
Then $V_\eff (\theta_H)$ in the model is given by
\beqn
&&\hskip -1.cm
V_\eff (\theta_H) \simeq 
4 I\Big[ \frac{1}{2} Q_0(q; \onehalf), \tilde \Lambda_{\rm gh} \Big] 
+ 2 I\Big[ \frac{1}{2 \cos^2 \theta_W} Q_0(q; \onehalf), \tilde \Lambda_{\rm gh} \Big] 
\cr
\noalign{\kern 10pt}
&&\hskip  .8cm
+ 3 I\Big[  Q_0(q; \onehalf), \tilde \Lambda_{\rm gh} \Big] 
- 12 \, I\Big[ \frac{1}{2(1+r_t)} Q_0(q; c_t), \tilde \Lambda_{\rm stop} \Big]
\label{effV5}
\eeqn
where $r_t=(m_b/m_t)^2$, and $\theta_W$ and $c_t$ are the Weinberg angle
and the bulk mass parameter for the top multiplet, respectively.  
The $\theta_H$-dependence enters through $Q_0(q;c)$.  
The values of the parameters with a given $z_L$ are summarized in Table \ref{parameters}.
The effective potential has the global minima at $\theta_H = \pm \onehalf \pi$ 
for $z_L \gg 1$.

\begin{table}[bht]
\begin{center}
\caption{The values of the parameters in the model employed in the evaluation of 
$V_\eff$ and $m_h$. The bulk mass parameter $c_t$ is determined from 
$m_t = 171.17\,$GeV. $\sin^2 \theta_W$ is determined by global fit of 
the forward-backword asymmetries in $e^+ e^-$ collisions on the $Z$ pole 
and the branching fractions of $Z$ decay.\cite{HTU2}
 $k$ and $m_\KK$ are in units of GeV.}
\label{parameters}
\vskip 5pt
\renewcommand{\arraystretch}{1.1}
\begin{tabular}{|c||c|c|c|c|}
\hline
$z_L$ 
& $k$
& ~$m_\KK$~
& ~$\sin^2 \theta_W$~
& $c_t$
\\
\hline
~$10^{15}$~ 
&
~$4.67 \times 10^{17}$~
&
1466
& 
0.2309
&
~0.432~
\\
\hline
$10^{17}$ 
&
$4.97  \times 10^{19}$
&
1562
& 
0.2310
&
0.440
\\
\hline
\end{tabular}
\end{center}
\end{table}

\begin{figure}[htb]
\begin{center}
\includegraphics[width=9cm]{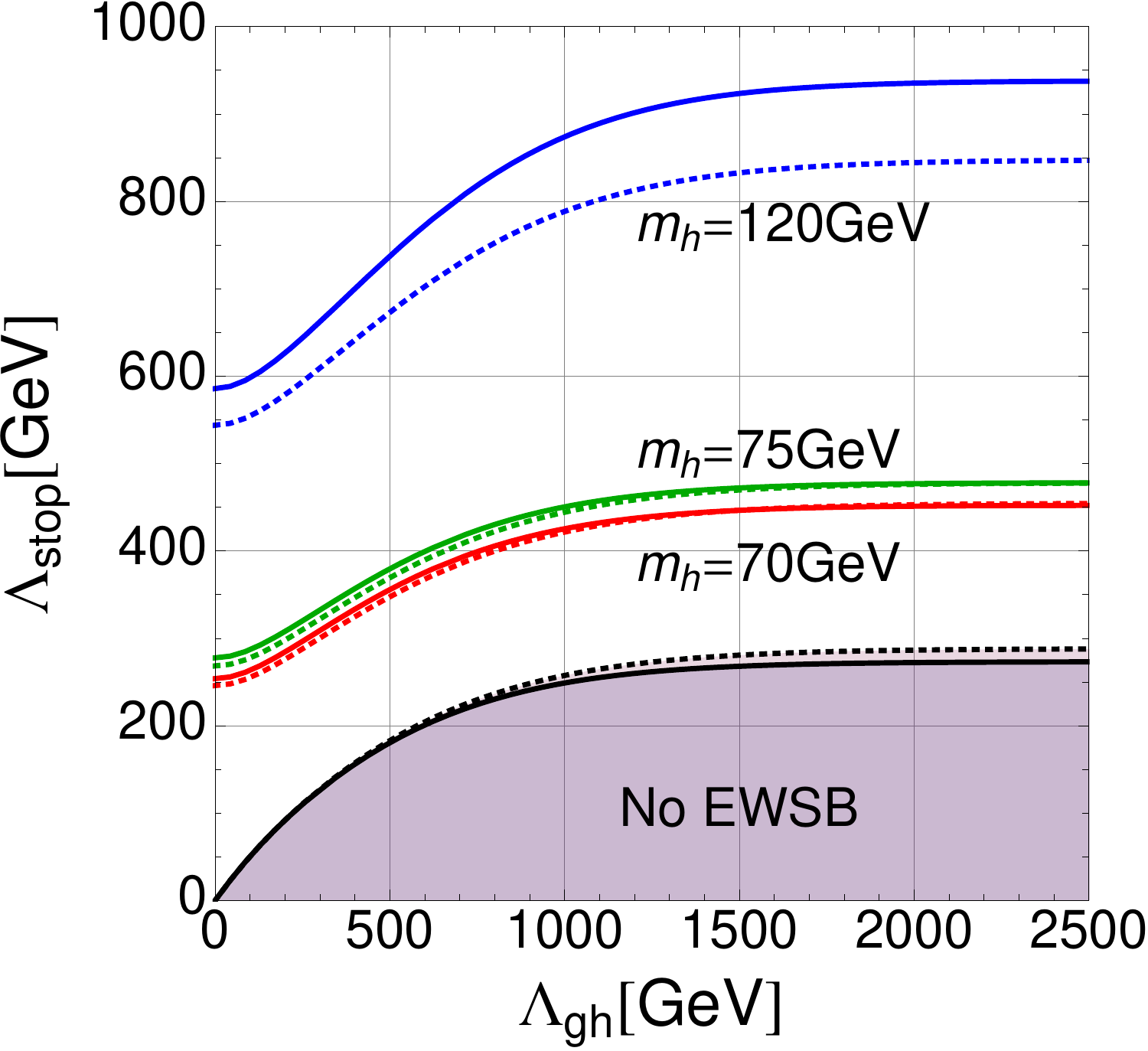}
\caption{SUSY breaking scales. Solid (dashed) lines correspond to $z_L = 10^{15}$
($10^{17}$).  Below the bottom solid (dashed) line the global minimum of
 $V_\eff(\theta_H)$ is located at $\theta_H =0$ so that the EW symmetry remains
unbroken.}
\label{susyB}
\end{center}
\end{figure}

The mass of the Higgs boson $m_h$ is related to the effective potential  
$V_\eff (\theta_H)$ by
$m_h^2 = f_H^{-2} (d^2 V_\eff / d \theta_H^2)|_{\theta_H =  \pi /2}$
where  $\onehalf g_w f_H = (k/L)^{1/2} (z_L^2 -1)^{-1/2} \sim m_W$.
$g_w$ is the 4D weak $SU(2)_L$ gauge coupling. 
Noting that the KK mass scale is given by $m_\KK \sim \pi k z_L^{-1}$, one finds
\beqn
&&\hskip -1.cm
m_h^2 = \frac{g_w^2 kL m_\KK^2}{32 \pi^4} \bigg\{ 
-4 J\Big[ \frac{1}{2} \Qbar_0(q; \onehalf), \tilde \Lambda_{\rm gh} \Big] 
- 2 J\Big[ \frac{1}{2 \cos^2 \theta_W} \Qbar_0(q; \onehalf), \tilde \Lambda_{\rm gh} \Big] 
\cr
\noalign{\kern 10pt}
&&\hskip  1.5cm
- 3 J\Big[  \Qbar_0(q; \onehalf), \tilde \Lambda_{\rm gh} \Big] 
+ 12 \, J\Big[ \frac{1}{2(1+r_t)} \Qbar_0(q; c_t), \tilde \Lambda_{\rm stop} \Big] \bigg\} ~,
\cr
\noalign{\kern 10pt}
&&\hskip -1.0cm
J[ f, \tilde \Lambda]   = \int_0^\infty dq \, q^3
\bigg\{ \frac{1}{1 + f[(q^2 + \tilde \Lambda^2)^{1/2}]} - \frac{1}{1 + f( q)} \bigg\} ~.
\label{Hmass}
\eeqn 
Given $z_L$, $m_h$ is determined as a function of $\Lambda_{\rm gh}$ and
$\Lambda_{\rm stop}$.  The result is summarized in fig.~\ref{susyB}.

For small $\Lambda_{\rm stop}$ ($\siml 200\,$GeV) 
with $\Lambda_{\rm gh} \simg 600\,$GeV,  $V_\eff (\theta_H)$ is minimized 
at $\theta_H = 0$ so that the EW symmetry remains unbroken.  
If both $\Lambda_{\rm gh}$ and $\Lambda_{\rm stop}$
are larger than 1$\,$TeV, the model is reduced to the non-supersymmetric model.

For $\Lambda_{\rm gh} \simg 1\,$TeV,  the desired $m_h = 70 \sim 75 \,$GeV is 
obtained with $\Lambda_{\rm stop} = 450 \sim 475 \,$GeV with tiny dependence
on $z_L$ in the range $10^{15} \sim 10^{17}$.   
With these values of $\Lambda_{\rm stop}$ one finds the mass of the stop to be
$m_{\tilde t} = 480 \sim 505 \,$GeV.  
In the analysis we have not specified masses of the sfermions except for the stop. 
If these sfermions are sufficiently heavy, evading the current bounds by LHC data, 
the stop or gravitino would become the lightest SUSY particle.

For $\Lambda_{\rm gh} \siml 100\,$GeV implying a light neutralino,  
the Higgs mass $m_h = 70 \sim 75 \,$GeV is 
obtained with $\Lambda_{\rm stop} = 250 \sim 275 \,$GeV, corresponding to
$m_{\tilde t} = 300 \sim 320 \,$GeV.
We stress that SUSY breaking scales for other quarks and
leptons can be much larger($\simg 1\,$TeV),  which does not affect the above
result.  There arises no constraint for gluino masses from this analysis so that
gluinos can be heavier than 1$\,$TeV.
$m_{\tilde t} = 300 \sim 320 \,$GeV with $\Lambda_{\rm gh} \siml 100\,$GeV
is in the range  allowed  by current experiments.\cite{LHCbound}  

In the above analysis we have supposed that the stop masses are degenerate. 
Though unnecessary in the current scheme, it may be of interest to see the 
effect of  large stop mixing, which plays an important role 
in MSSM to obtain a desired Higgs mass\cite{Higgs125}.
In the presence of the left-right squark mixing,  the stop masses become non-degenerate.
The spectra of their KK towers are approximated by  
$m_n^{\text{stop},i} = \sqrt{(m_n^{\text{top}})^2+\Lambda_{\text{sotp},i}^2}$, 
$i=1,2$, and accordingly the last terms in Eq.~(\ref{effV5}) and in the first equation 
of (\ref{Hmass}) are separated into two parts. As an extremal case we  consider the case where
one of the stops is very heavy and decouple.  In such a case the curves
in Fig.~\ref{susyB} are shifted downward. 
For example we obtain $\Lambda_{stop} \sim 260$~GeV [$600$~GeV] for 
$\Lambda_{\text{gh}}= 100$~GeV [$1000$~GeV] to obtain $m_h=120$~GeV. 
The Higgs mass $m_h$ can be lowered only to 110GeV [84GeV] for 
$\Lambda_{\rm gh} = 100$~GeV [1000 GeV]. 
To obtain $m_h = 70 \sim75\,$GeV, it is desirable to have approximately degenerate 
stop masses in the current scheme.

In the present analysis we adopted a mass spectrum of a SUSY KK partner 
in the form (\ref{spectrum1}) for convenience.  Depending on how SUSY is broken,
the spectrum may deviate from  (\ref{spectrum1}).  However, the detailed form
of the spectrum is not relevant in the present analysis, provided that 
$\mbar_n > m_n$ and $\mbar_n \sim m_n$ for $m_n \gg \Lambda$.
Only low lying modes in the KK towers give relevant contributions to 
the $\theta_H$-dependent part of $V_\eff (\theta_H)$. 
Contributions coming from the modes with $m_n, \mbar_n \gg m_\KK$ 
are irrelevant. 

We need a further consideration of the consistency with the current
electroweak precision measurements, especially with Peskin-Takeuchi $S$ 
and $T$ parameters.\cite{PeskinTakeuchi}  
There are many studies on these parameter in the models of 
extra dimensions.\cite{Giudice2007, Cacciapaglia1, Agashe3}
The composite Higgs models, which are regarded as holographic duals of the 
five-dimensional gauge-Higgs unification models, are severely constrained by the 
precision measurements.
On the other hand it has been shown
that the gauge couplings of gauge bosons, leptons
and quarks in the $SO(5) \times U(1)$ gauge-Higgs unification model 
deviate little from those in the standard model,\cite{HNU, HTU2}
which indicates subtle difference between the composite Higgs models
and the $SO(5) \times U(1)$ gauge-Higgs unification model.  
The spontaneous breaking of $SO(4) \times U(1)$ to $SU(2) \times U(1)$ 
triggered by a brane scalar field is crucial to have a realistic model of
the electroweak symmetry breaking, which has
not been properly taken into account in the literature.
Recently it has been noticed that the symmetry group of the standard model
may rotate in the $SO(5)$ group space according to  the value of 
$\theta_H$.\cite{Contino2011} This certainly  necessitates reexamination
of $S$ and $T$ in the model.  

To conclude,  the dark matter of the Universe can be explained
by stable Higgs bosons in the supersymmetric extension of the 
$SO(5) \times U(1)$ gauge-Higgs unification in the RS spacetime
without spoiling the consistency with collider data at low energies, 
if $m_{\tilde t} = 300 \sim 320 \,$GeV  when
gauginos in the electroweak sector  are light.
The masses of gluinos as well as  other squarks and sleptons 
do not affect the result.
It would be of extreme importance to find  the stop $\tilde t$ at LHC
to get insight into the structure of spacetime.

Besides studying  the electroweak  precision observables, 
it is necessary  to complete the model by incorporating flavor  
mixings\cite{Lim3} and implementing light neutrinos by the 
seesaw mechanism in the bulk-brane system.
We hope to report on these in the near future.

\vskip 10pt

{\bf Acknowledgements:} 
The authors would like to thank K.\ Oda  and M.\ Tanaka for valuable comments.
This work was supported in part 
by  scientific grants from the Ministry of Education and Science, 
Grants No.\ 20244028, No.\ 23104009 and  No.\ 21244036.

\vskip 1cm

\renewenvironment{thebibliography}[1]
         {\begin{list}{[$\,$\arabic{enumi}$\,$]}  
         {\usecounter{enumi}\setlength{\parsep}{0pt}
          \setlength{\itemsep}{0pt}  \renewcommand{\baselinestretch}{1.2}
          \settowidth
         {\labelwidth}{#1 ~ ~}\sloppy}}{\end{list}}

\def\jnl#1#2#3#4{{#1}{\bf #2},  #3 (#4)}

\def\Zphys{{\em Z.\ Phys.} }
\def\jssc{{\em J.\ Solid State Chem.\ }}
\def\jpsJ{{\em J.\ Phys.\ Soc.\ Japan }}
\def\ptps{{\em Prog.\ Theoret.\ Phys.\ Suppl.\ }}
\def\PTP{{\em Prog.\ Theoret.\ Phys.\  }}
\def\JMP{{\em J. Math.\ Phys.} }
\def\NPB{{\em Nucl.\ Phys.} B}
\def\NP{{\em Nucl.\ Phys.} }
\def\PLB{{\it Phys.\ Lett.} B}
\def\PL{{\em Phys.\ Lett.} }
\def\PRL{\em Phys.\ Rev.\ Lett. }
\def\PRB{{\em Phys.\ Rev.} B}
\def\PRD{{\em Phys.\ Rev.} D}
\def\PRe{{\em Phys.\ Rep.} }
\def\AP{{\em Ann.\ Phys.\ (N.Y.)} }
\def\RMP{{\em Rev.\ Mod.\ Phys.} }
\def\ZPC{{\em Z.\ Phys.} C}
\def\SCI{\em Science}
\def\CMP{\em Comm.\ Math.\ Phys. }
\def\MPLA{{\em Mod.\ Phys.\ Lett.} A}
\def\IJMPA{{\em Int.\ J.\ Mod.\ Phys.} A}
\def\IJMPB{{\em Int.\ J.\ Mod.\ Phys.} B}
\def\EPJC{{\em Eur.\ Phys.\ J.} C}
\def\PR{{\em Phys.\ Rev.} }
\def\JHEP{{\em JHEP} }
\def\JCAP{{\em JCAP} }
\def\cmp{{\em Com.\ Math.\ Phys.}}
\def\JPA{{\em J.\  Phys.} A}
\def\JPG{{\em J.\  Phys.} G}
\def\NJP{{\em New.\ J.\  Phys.} }
\def\CQG{\em Class.\ Quant.\ Grav. }
\def\ATMP{{\em Adv.\ Theoret.\ Math.\ Phys.} }
\def\ibid{{\em ibid.} }

\renewenvironment{thebibliography}[1]
         {\begin{list}{[$\,$\arabic{enumi}$\,$]}  
         {\usecounter{enumi}\setlength{\parsep}{0pt}
          \setlength{\itemsep}{0pt}  \renewcommand{\baselinestretch}{1.2}
          \settowidth
         {\labelwidth}{#1 ~ ~}\sloppy}}{\end{list}}

\def\reftitle#1{}                
\def\archive#1{}               


\end{document}